\documentclass[12pt]{article}

\usepackage{amssymb}

\begin{document}

  \def\be{\begin{equation}}
  \def\ee{\end{equation}}
  \def\bea{\begin{eqnarray}}
  \def\eea{\end{eqnarray}}
  \def\nn{\nonumber}
  \def\l{\lambda}
  \def\t{\times}
  \def\[{\bigl[}
  \def\]{\bigr]}
  \def\({\bigl(}
  \def\){\bigr)}
  \def\p{\partial}
  \def\o{\over}
  \def\ta{\tau}
  \def\cm{\cal M}
  \def\R{\bf R}
  \def\b{\beta}
  \def\a{\alpha}
\newcommand{\WP}{\mathbf{WP}}
\newcommand{\la}{\lambda}

\baselineskip=18.6pt plus 0.2pt minus 0.1pt

\begin{titlepage}
\title{
\begin{flushright}
{\normalsize GNPHE/05-15}\\ \mbox{}
\end{flushright}
{\bf On ADE Quiver  Models}
\\[.3cm]
{\bf and  F-Theory Compactification  } }

\author{A. Belhaj$^{1,}$\thanks{{\tt abelhaj@uottawa.ca}}\ ,\ \
J. Rasmussen$^{2,}$\thanks{{\tt J.Rasmussen@ms.unimelb.edu.au}}\ ,\
\ A. Sebbar$^{1,}$\thanks{{\tt sebbar@mathstat.uottawa.ca}}\ ,\ \
M.B. Sedra$^{3,4,}$\thanks{{\tt sedra@ictp.it}}
\\[8pt]
{\it \small $^1$ Department of Mathematics and Statistics,
University of Ottawa}\\{ \it \small 585 King Edward Ave., Ottawa,
ON, Canada,  K1N 6N5  }\\ {\it \small $^2$ Department of
Mathematics and Statistics, University of Melbourne }\\{ \it
\small Parkville, Victoria 3010, Australia}\\ {\it \small $^3$
Laboratoire de Physique de la Mati\`{e}re et Rayonnement (LPMR),
K\'{e}nitra, Morocco }\\ { \it \small   Facult\'e des Sciences,
Universit\'e Ibn Tofail, Kenitra, Maroc }\\ { \it \small $^4$
Virtual African Center for Basic Sciences and Technology
(VACBT)}\\{ \it \small Groupement National de Physique des Hautes
Energies (GNPHE)}\\{\it \small  Facult\'e des Sciences, Rabat,
Maroc }\\ } \maketitle

\begin{abstract}
Based on mirror symmetry, we discuss geometric engineering of $N=1$
$ADE$ quiver models
from F-theory compactifications on elliptic K3 surfaces fibered
over certain four-dimensional base spaces.
The latter are constructed as intersecting 4-cycles according
to $ADE$ Dynkin diagrams, thereby mimicking the construction
of Calabi-Yau threefolds used in geometric engineering in
type II superstring theory. Matter is incorporated by considering
D7-branes wrapping these 4-cycles. Using a
geometric procedure referred to as folding, we discuss how the
corresponding physics can be converted into a scenario with
D5-branes wrapping 2-cycles of ALE spaces.
\end{abstract}

\end{titlepage}

\newpage

\section{Introduction}

The construction of four-dimensional supersymmetric gauge theories
has attracted much attention and been investigated from various
points of view in superstring compactifications, for example. The
approach of interest here is the construction of gauge theories from
geometric data of superstring backgrounds. Thus, the gauge group and
matter content of the resulting models are obtained from the
singularities of the K3 fibers and the non-trivial geometry
describing the base space of the internal manifolds. In this way,
the complete set of physical parameters of the gauge theory is
related to the moduli space of the associated manifolds. This
program is called geometric engineering \cite{KKV,
KMV,BFS,ABS2,ABS1,BS1}. It enables one to represent supersymmetric
gauge models by quiver diagrams similar to Dynkin graphs of
ordinary, affine or indefinite Lie algebras \cite{KMV, ABS2, ABS1}.
One considers the base space to be composed of a collection of
intersecting  compact cycles, each of which is giving rise to an
$SU$ gauge group factor. To each of these is associated a Dynkin
node, and for each pair of groups with matter in bi-fundamental
representations, the two corresponding nodes are connected by a
line.

Four-dimensional  quiver gauge models  have been constructed also in
M-theory on  seven-dimensional manifolds with $G_2$  holonomy. The
compactification manifolds are K3 fibrations over  three-dimensional
base spaces with $ADE$ geometries. The resulting gauge theories have
been discussed in the realm of (p,q) brane webs \cite{BDR}.

The aim of the present paper is to contribute to the program of
geometric engineering by constructing four-dimensional $N=1$ $ADE$
quiver models from F-theory compactifications. The manifolds are
elliptic K3 surfaces fibered over $ADE$  4-cycles, where the base is
obtained by solving the $ADE$ hyper-K\"ahler singularities
\cite{BS2}, and are similar to the Calabi-Yau threefolds used in
type II geometric engineering \cite{KKV}. An objective here is to
construct explicit  models  of such geometries  leading to gauge
theories with bi-fundamental  matter in four dimensions. As an
illustration, we consider $A_r$ quiver gauge  models by introducing
4-cycles  in the base which are intersecting according to an $A_r$
Dynkin graph. In particular, we consider in some details the cases
of $A_1$ and $A_2$, and we find that they are linked to ordinary
$A_1$ and $A_2$ singularities of the asymptotically locally
Euclidean (ALE) spaces. The dual type IIB models involve  D7-branes
wrapping  $ADE$ 4-cycles.  Using a geometric procedure referred to
as folding, we show that the corresponding physics  can be converted
into a scenario with D5-branes wrapping 2-cycles  of  ALE spaces.

The present paper is organized as follows. Section 2
provides a brief review on how
geometric engineering may be used to obtain
four-dimensional gauge models from superstring theory or M-theory.
The basic extension to F-theory is discussed in Section 3
where focus is on the construction of $ADE$ fourfolds.
Mirror symmetry is employed in Section 4 when studying the
resulting physics in the presence of D7-branes.
The folding procedure linking this to a similar
study of D5-branes in superstring theory is also discussed
in Section 4. Section 5 contains some concluding remarks.

\section{Geometric engineering}

In this section, we review   briefly  the main steps in obtaining
four-dimensional  supersymmetric  gauge models either from
compactification of type II superstrings on Calabi-Yau threefolds
$CY^3$ \cite{KKV, BFS}, or from compactification of M-theory on
$G_2$ manifolds \cite{BDR}. In either case, the manifold is a K3
fibration over a base space $B$.  The basic tasks are to specify the
singularity of the K3 fibration manifolds and to take the limit in
which the volume $V(B)$ of the base space is very large so that
gravitational effects may be ignored. In particular, one considers
the K3 fibers locally as non-compact ALE spaces with $ADE$
singularities. One subsequently examines the compactification  in
the presence of D2-branes or M2-branes wrapping the vanishing
2-cycles in the ALE spaces. This enables one to make conclusions
about the gauge group $G$ and matter content in the four-dimensional
model. This analysis can be carried out in two steps as one may
perform an initial but partial compactification on the K3 surfaces
followed by a further compactification down to four dimensions.

To illustrate this, we consider type IIA superstring theory. Let us
study the  simplest case of  $SU(2)$ gauge  theory  obtained from an
$A_1$ singularity of the ALE fiber space at the origin.
Mathematically,  this is described by  \be
xy+z^2=0
\label{xyz2}
\ee
where $x$, $y$ and $z$ are complex coordinates. As usual, the
singularity can be removed either by deforming the complex
structure or by a blow-up procedure. Geometrically, this
corresponds to replacing the singular point $(x=y=z=0)$ by a
$\mathbf{CP}^{1}\sim\mathbf{S}^{2} $.

In the case where the $A_1$ singularity has been resolved, a
D2-brane  can wrap around the blown-up $\mathbf{CP}^{1}$ giving rise
to a pair of vector particles, $W^{\pm}$. These particles correspond
to the two possible orientations of the wrappings. The particles
have masses proportional to the volume of the blown-up two-spheres.
$W^{\pm}$ are charged under the $U(1)$ field $Z$ obtained by
decomposing the type IIA superstring three-form $C_{\mu\nu\lambda}$
in terms of the harmonic two-form on the two-sphere and a one-form
not in the K3 fiber. In the  singular limit where $\mathbf{CP}^{1}$
shrinks to a point, the three vector particles become massless, and
they form an adjoint $SU(2)$ representation. We thus obtain an $N=2$
$SU(2)$ gauge symmetry in six dimensions.

A further compactification on the base space $B_2$ gives pure
$SU(2)$ Yang-Mills theory in four dimensions. $N=2$  models are
obtained by  taking $B_2$  as a real two-sphere. If we instead
compactify on Riemann surfaces with one-cycles, we obtain
$N=4$ gauge models. The extra scalar fields that one can get in such an
$N=4$ system can be identified with the expectation values of Wilson
lines on  such a surface.

Before turning to geometric engineering in
F-theory, we list a couple of comments on extensions of the superstring
compactification indicated above:
\begin{itemize}
\item
The geometric  analysis of  the $SU(2)$  model   can be extended
straightforwardly to general simply-laced $ADE$ gauge groups. This
extension is based on the classification of  $ADE$ singularities of
ALE spaces.
\item Matter is  incorporated by considering intersecting
2-cycles in the base space $B_2$. In this case, the base is obtained
by solving $ADE$ singularities  of ALE spaces \cite{KMV}.
\item There are three kinds of models and they are related to the
classification of
generalized Cartan matrices of Kac-Moody algebras \cite{ABS1,ABS2}.
\item   Similar comments apply to the case of M-theory on manifolds
with $G_2$ holonomy in which case the D2-branes are replaced by
M2-branes \cite{BDR}.
\end{itemize}
In what follows, we study geometric engineering in F-theory by
introducing intersecting geometries in the accompanying
compactifications on Calabi-Yau fourfolds. These manifolds  are
constructed as elliptic K3 surfaces fibered over intersecting
4-cycles according to $ADE$  Dynkin graphs. This geometric
engineering may result in $N=1$ $ADE$ quiver models with
bi-fundamental matter in four dimensions, thus extending the result of 
\cite{Vafa1}.

\section{Geometric engineering in F-theory  compactification}

In this section, we study geometric engineering of quiver gauge
models in F-theory compactification. It is recalled that F-theory
defines a non-perturbative vacuum of type IIB superstring theory in
which the dilaton and axion fields of the superstring theory are
considered dynamical. This introduces an extra complex modulus which
is interpreted as the complex parameter of an elliptic curve thereby
introducing a non-perturbative vacuum of the type IIB superstring in
a {\em twelve-dimensional} space-time \cite{Vafa2}. F-theory may
also be defined in the context of string dualities, and as we will
discuss, F-theory on elliptically fibered Calabi-Yau manifolds may
be understood in terms of dual superstring models.

It is also recalled that type IIB superstring theory is a
ten-dimensional model of closed strings with chiral $N=2$
supersymmetry. The bosonic fields of the corresponding low-energy
field theory are the graviton $g_{\mu\nu}$, the anti-symmetric
tensor $B_{\mu\nu}$ and the dilaton $\phi$ coming from the NS-NS
sector and the axion $\xi$, and the anti-symmetric tensor fields
$\tilde B_{\mu\nu}$ and the self-dual four-form
$D_{\mu\nu\rho\sigma}$ stemming from the R-R sector. There is no
non-abelian gauge field in the massless spectrum of type IIB
superstring theory which instead contains D$p$-branes with
$p=-1,1,3,5,7$ and  $9$ on which the gauge fields $A_{\mu}$ live.
These extended objects are non-perturbative solutions playing a
crucial role in the study of gauge theories in superstring models.
It is also noted that type IIB superstring theory has a
non-perturbative $PSL(2,\mathbb{Z})$ symmetry with respect to which
the fields $g_{\mu\nu}$ and $D_{\mu\nu\rho\sigma}$ are invariant.
The complex string coupling $\tau_{IIB}=\xi+ie^{-\phi}$ and the
doublet $(B_{\mu\nu},\tilde B_{\mu\nu})$ of two-forms, on the other
hand, are believed to transform as
\be
\tau_{IIB} \to {a\tau_{IIB}+b\over c\tau_{IIB}+d},\ \ \ \ \ \ \  a,b,c,d \in
\mathbb{Z},
\ee
and
\be
{B_{\mu\nu}\choose \tilde B_{\mu\nu}}\to
\left(\matrix{ a&b\cr c&d\cr}\right){B_{\mu\nu}\choose \tilde
B_{\mu\nu}},
\ee
where the integers $a,b,c$ and $d$ satisfy $ab-cd=1$.

Following Vafa \cite{Vafa2},
one may interpret the complex field $\tau_{IIB}$ as
the complex structure of an extra torus $T^2$ resulting in the
aforementioned twelve-dimensional model. From this point of view,
type IIB superstring theory may be seen as the compactification of
F-theory on $T^2$. Starting from F-theory, one can similarly look
for new superstring models in lower dimensions obtained by
compactifications on elliptically fibered Calabi-Yau manifolds. For
example, the eight-dimensional F-theory on elliptically fibered K3
is obtained by taking  a two-dimensional complex  compact manifold
given by \be
y^2=x^3+f(z)x+g(z),
\ee where $f$ and $g$ are polynomials of degree 8 and 12,
respectively. One varies the $\tau$ torus over the points of a
compact space which is taken to be a Riemann sphere $\mathbf{CP}^1$
parametrized by the local coordinate $z$. In other words, the
two-torus complex structure $\tau (z)$ is now a function of $z$ as
it varies over the $\mathbf{CP}^1$ base of K3. The above compact
manifold generically has 24 singular points corresponding to $\tau
(z)\to \infty$. These singularities have a remarkable physical
interpretation as each one of the 24 points is associated with the
location of a D7-brane in non-perturbative type IIB superstring
theory.  We assume that the number of D7-branes is arbitrary for
non-compact two-dimensional complex space.

We now turn to the study of geometric engineering of quiver gauge
models in F-theory. The method we will be using here is similar to
the one employed in the engineering of four-dimensional QFT$_4$ from
type II superstring theory or M-theory on $G_2$ manifolds. Thus, we
first build the local fourfolds enabling us to construct
four-dimensional quiver gauge models with bi-fundamental matter from
compactification of F-theory. The manifolds we will consider are
quite simple and involve K3 fibrations specifying the gauge groups
in four dimensions.  As in the study of string theory, matter may be
incorporated by introducing a non-trivial geometry in the base
space, and we are naturally led to consider a base space involving a
collection of intersecting 4-cycles. Each 4-cycle gives rise to a
gauge-group factor while matter is described in terms of
bi-fundamental representations of these gauge groups.

\subsection{Construction of fourfolds}

The manifolds that we propose to use can be described as
hyper-K\"ahler quotients as they are of the form \be
\label{fourfold}
X_8: \quad C^2\,\,\mbox{fibered over}\,\, V^2.
\ee
Here, $V^2$ is a two-dimensional toric variety specified below,
while $C^2$ is the fiber which can  be converted into a local elliptic
K3 surface  by orbifold actions. The manifold $X_8$ thus has 8 real
dimensions. An advantage of working with a geometry like
(\ref{fourfold}) is that one may study an intersecting structure in
the base $V^2$ using physical arguments. This is done in part by
mimicking the similar analysis of $ADE$ singularities in terms of
$N=2$ toric sigma models. In particular, the manifold can appear in
the construction of a two-dimensional field theory  with $N=4$
supersymmetry. This model has $r+2$  hypermultiplets  which are
charged  under $r$ abelian vector multiplets  $U(1)^r$ with charges
$Q^a_i$. The geometry given by (\ref{fourfold}) solves the D-flat
constraint equations of this $N=4$ field theory: \be \label{D-terms}
    \sum\limits_{i=1}^{r+2} Q^a_i\[\phi_i^{\alpha}{\bar \phi}_{i \beta}
           +\phi_{i \beta}
          {\bar \phi}_i^{\alpha}\]=\vec \xi_a
          \vec\sigma^\alpha_\beta, \quad\ \ \   a=1,\ldots,r
\ee where the $\phi_i^{\alpha}$'s denote $(r+2)$-component complex
doublets of hypermultiplets. In this equation,  $\vec \xi_a$ is
a 3-vector coupling parameters  rotated by $SU(2)$ symmetry, and
$\{\vec \sigma^\alpha_\beta\}$ are the traceless $2\t2$ Pauli matrices.
Up to some details, (\ref{D-terms}) can be rewritten as follows \bea
\sum\limits_{i=1}^k Q_i^a(
  |\phi^1_i|^2-|\phi^2_i|^2) &= &\xi^3 _a,\\
\sum\limits_{i=1}^kQ_i^a
  \phi^1_i {\bar \phi}_{i}^2&=&\xi^1_a+i\xi^2_a,\\
\sum\limits_{i=1}^k Q_i^a\phi^2_i
  {\bar \phi}_{i}^1&=&\xi^1_a-i\xi^2_a.
\eea Notice that for  $\phi^2_i=0$ and $\xi^{1,2}_a=0$,
(\ref{D-terms}) describes the two-dimensional toric base $V^2$,
which in the $N=2$  sigma model analysis   is  given by \be
  \sum\limits_{i=1}^{r+2} Q_i^a|\phi^1_i|^2 = \xi^3 _a, \qquad
a=1,\ldots,r.
\ee It follows that $\phi^1_i$ can be taken as toric  coordinates of
$V^2$, while $\phi^2_i$ can be regarded  as coordinates of the fiber
$ C^2$.

In this way, the intersection matrix  of the 4-cycles defining the
base can be identified with the charge matrix of an $N=4$ gauge
theory in two dimensions. As we will see, this matrix can take the
form of a Cartan matrix of an $ADE$ Lie algebra. Geometric
engineering in F-theory compactification thereby enables us to
describe four-dimensional $ADE$ quiver gauge models with
bi-fundamental matter.

\subsection {$ADE$  fourfolds}

As already indicated, we will consider fourfolds whose base spaces
are constructed as 4-cycles intersecting according to $ADE$ Dynkin
diagrams. We refer to these manifolds as $ADE$ fourfolds. They
constitute a very natural class of manifolds in this context. It is
pointed out, though, that we in principle could consider more
complicated geometries. We will restrict ourselves to the $ADE$
fourfolds as they allow us to extract the corresponding physics in a
straightforward manner.

For simplicity, we will mainly consider the case of $A_r$. In
two-dimensional $N=4$ field theory, the matrix $ Q_i^a$ can then be
identified with the Cartan matrix of  $ A_r$  Lie algebras. The
charge matrix thus reads
\be
Q^a_i=\delta^a_{i}-2\delta^{a+1}_{i}+\delta^{a+2}_{i}, \ \ \ \ \ \ \ \ \
  a=1,\ldots,r,\ \ \ \ \ i=1,\ldots,r+2.
\ee Inserting this matrix  into the D-flatness  condition
(\ref{D-terms}), one obtains the following system of $3r$ equations:
\bea \label{ADE}
   (|\phi^1_{a-1}|^2+|\phi^1_{a+1}|^2-2|\phi^1_{a}|^2)-(|\phi^2_{a-1}|
   ^2+|\phi^2_{a+1}|^2-
   2|\phi^2_a|^2)&=&\xi_a, \\
  \phi^1_{a-1}{\bar \phi}^2_{a-1}+\phi^1_{a+1}{\bar \phi}^2
  _{a+1}-2\phi^1_{a} {\bar \phi}^2_{a}&=&0,\\
  \phi^2_{a-1}{\bar \phi}^1_{a-1}+\phi^2_{a+1}{\bar \phi}^1
  _{a+1}-2\phi^2_{a} {\bar \phi}^1_{a}&=&0 .
\eea A simple examination  of these equations reveals   that    the
base consists of  $r$ intersecting $\mathbf{WP}^2_{1,2,1}$ according
to the following  $A_r$  Dynkin diagram \be
    \mbox{
         \begin{picture}(20,30)(70,0)
        \unitlength=2cm
        \thicklines
    \put(0,0.2){\circle{.2}}
     \put(.1,0.2){\line(1,0){.5}}
     \put(.7,0.2){\circle{.2}}
     \put(.8,0.2){\line(1,0){.5}}
     \put(1.4,0.2){\circle{.2}}
     \put(1.6,0.2){$.\ .\ .\ .\ .\ .$}
     \put(2.5,0.2){\circle{.2}}
     \put(2.6,0.2){\line(1,0){.5}}
     \put(3.2,0.2){\circle{.2}}
     \put(-1.2,.15){${\bf A_{r}:}$}
   \end{picture}
} \label{ordAk} \ee

As an illustration, we now consider the $A_1$ geometry corresponding
to setting $r=1$ in (\ref{ADE}). Its description involves    three
hypermultiplets $\phi_i$ and one charge vector $Q_i= (1,-2,1)$.
After permuting the roles of $\phi^1_2$ and ${\bar \phi}^2_{2}$  and
making the following  field changes $\phi^1_{1}=\varphi_1$,
$\phi^2_{1}=\psi_1$ $\phi^1_{3}=\varphi_2$, $\phi^2_{3}=\psi_3$, $ -
{\bar \phi}^2_{2}=\varphi_2$, ${\bar \phi}^1_{2}=\psi_2$, there are
only  three constraint equations, namely \bea
   (|\varphi_{1}|^2+|\varphi_{3}|^2+2|\varphi_{2}|^2)-(|\psi_{1}
   |^2+|\psi_{3}|^2+2|\psi_2|^2) &=&\xi^3, \\
  \varphi_{1}{\bar \psi}_{1}+\varphi_{3}
  {\bar \psi}_{3}+2\varphi_{2}{\bar \psi}_{2}&=&0,\\
  {\bar \varphi}_{1} {\psi_{1}}+
  {\bar \varphi}_{3}\psi_{3}+2{\bar \varphi}_{2}\psi_{2}&=&0.
\eea It is noted that for  $\psi_{1}=\psi_{2}=\psi_3=0$, these
equations reduce to
$ |\varphi_{1}|^2+|\varphi_{3}|^2+2|\varphi_{2}|^2=\xi^3$
describing $\mathbf{WP}^2_{1,2,1}$
with area proportional to $\xi$. Thus, the total  geometry  is  now
given by  $C^2\times \mathbf{WP}^2_{1,2,1}$.  As already indicated,
for generic values of $r$, $V^2$ can be identified with a system of
intersecting $\mathbf{WP}^2_{1,2,1}$ according to the $A_r$ Dynkin
diagram (\ref{ordAk}). In this case, $ \mathbf{WP}^2_{1,2,1}$ is
playing the role of $ \mathbf{CP}^1$ in the ALE spaces.

\section{$ADE$ quiver gauge models in F-theory compactification}

Having constructed a class of $ADE$ fourfolds, we now discuss the
physics resulting from compactifying F-theory on these manifolds.
Our analysis  here is  based  on a dual type IIB superstring
description where the compactification on Calabi-Yau threefolds is
considered in the presence of D7-branes wrapping 4-cycles and
filling the four-dimensional  Minkowski space-time. As in the string
theory, we initially compactify on the K3 fiber. This results in an
eight-dimensional  supersymmetric gauge theory which can be
identified with a gauge model living in the world-volume of the
D7-branes. A subsequent compactification of F-theory to four
dimensions is then equivalent to wrapping the D7-branes on $V^2$ in
the type IIB superstring compactification. In this scenario, $V^2$
is taken to be embedded in the three-dimensional Calabi-Yau
compactification space.

The present task is therefore to look for the type IIB  geometries
dual to the F-theory on $ADE$ fourfolds.  Since $V^2$ is a compact
toric manifold, it is natural to expect  that  the type IIB
geometry may be  a toric variety as well.  At first sight, such a
manifold appears quite complicated. As we will see, however, if we
restrict ourselves to the physics  coming from the D7-branes, the
dual type IIB geometry may be described as a toric Calabi-Yau
manifold with $V^2$ as base space. We will assume that it
corresponds to a line bundle over $V^2$. It is also noted that the
toric property allows us to describe the corresponding D7-brane
physics in terms of the toric data of the Calabi-Yau threefold.

We are interested in fourfolds with $ADE$ intersecting 4-cycles in
the base, and we will discuss how F-theory on these $ADE$ fourfolds
can be interpreted in terms of type IIB superstrings on Calabi-Yau
threefolds in the presence of D5-branes wrapping $ADE$ intersecting
2-cycles of ALE spaces. This connection is based on a geometric
procedure called folding. For type IIB toric geometry, this
procedure has been used to geometrically engineer non-simply-laced
quiver gauge theories \cite{BFS}. In brief, one identifies the toric
vertices of the Calabi-Yau threefold which are permuted under the
folding action $\Gamma$ (which is an outer automorphism of the
associated toric graph). This imposes certain constraints on the
toric data depending on the precise action of $\Gamma$. It turns out
that such actions become very simple  using  local mirror
transformations and that there are two possible representations
\cite{BFS}. The one we are interested in here has the property of
resulting in geometries with one dimension less than the `natural'
one. This dimensional reduction follows straightforwardly from the
toric data of the resulting geometry. {}From the string theory point
of view, however, this missing complex direction will resurface.
This will be addressed below.

\subsection{$A_1$ quiver  model}

F-theory on an $A_1$ fourfold   is   expected to be equivalent to
type IIB superstring theory on ${\cal O}_{\mathbf{WP}^2_{1,1,2}}
(-4)$ in the presence  of D7-branes having a world-volume
$\mathbb{R}^4 \times \mathbf{WP}^2_{1,1,2}$.  This would give $N=1$
pure  Yang-Mills theory in four dimensions, and the Yang-Mills
coupling constant $g_{YM}$ is related to the volume of  $
\mathbf{WP}^2_{1,1,2}$ as \be
\mbox{Vol}( \mathbf{WP}^2_{1,1,2})=1/g_{YM}^2.
\ee

\subsection{Models with bi-fundamental matter}

Here, we discuss how to incorporate bi-fundamental  matter in
geometric engineering  of  F-theory. Such matter can be introduced
by replacing the single  $\mathbf{WP}^2_{1,1,2}$ considered above by
an intersecting  geometry according to $ADE$ toric Dynkin graphs. As
we will see, these complicated gauge theories can be reduced to
well-known models  based on ordinary $ADE$ singularities of ALE
spaces involving $\mathbf{CP}^1$ complex curves. In this analogy,
the D7-branes wrapping the intersecting $\WP_{1,1,2}^2$ constituents
are replaced by D5-branes wrapping the intersecting $\mathbf{CP}^1$
constituents of the deformed ALE spaces. This connection is based on
the geometric procedure of folding which we alluded to above. To
illustrate the basic mechanism, we initially consider the example or
toy model $A_1$.

We consider ${\cal O}_{\mathbf{WP}^2_{1,1,2}} (-4)$    which is a
toric manifold. It can be described by an $N=2$  gauged linear sigma
model with 4 chiral superfields and $U(1)$  gauge symmetry with
respect to which the superfields have  charges \be q_i=(1,1,2,-4)
\label{vecwp2} \ee satisfying the constraint \be
\sum\limits_{i=1}^4q_i=0. \ee This condition implies that the space
is a local Calabi-Yau manifold. In toric geometry,   it can be
represented by \be \label{toric1} \sum\limits_{i=1}^4q_iv_i=0, \ee
and a particular toric vertex  realization of this manifold   is
given by \be \label{vertex} v_1=(-2,1,1),\quad v_2=(2,1,1),\quad
v_3=(0,-1,1), \quad v_4=(0,0,1). \ee

\subsubsection{Mirrors  of ${\cal O}_{\mathbf{WP}^2_{1,1,2}} (-4)$}

As usual, the mirror  of ${\cal O}_{\mathbf{WP}^2_{1,1,2}} (-4)$ is
obtained by  solving  the following constraint equations
\cite{HV,HIV,A} \bea
\sum\limits_{i=1}^4a_iy_i&=&0,\nonumber\\
\prod\limits_{i=1}^4y_i^{Q_i}&=&1, \label{mirror}\eea where $\{a_i\}$
are the complex parameters defining the complex structure of the
mirror geometry. However, only one of these parameters is physical
and it describes the mirror of the size of ${\mathbf{WP}^2_{1,1,2}}$. For
simplicity, one can fix also this parameter to 1.

It turns out that there  are many ways to solve the
mirror  system (\ref{mirror}). A nice way is to introduce two complex
variables
$U_i=(U_1,U_2)$, specified later on, and a system of two-dimensional
vectors of integer entries: $\{n_i, i=1, \ldots,4\}$. A simple
examination reveals that (\ref{mirror}) can be solved   using the
following parametrization \be
y_i=\prod\limits_{j=1}^2U_j^{n_i^j},\quad i=1,2,3,4
\ee subject to \be
  \label{toric2}
\sum\limits_{i=1}^4 q_in_i^1=0,\ \ \ \ \ \ \ \sum\limits_{i=1}^4
q_in_i^2=0. \ee In this way, the corresponding mirror geometry reads
\be \sum\limits_{i=1}^4\prod\limits_{j=1}^2U_j^{n^j_i}=0. \ee
Setting $v_i=(n_i,1)$,  we get the following mirror geometry \be
\label{mgeometry} 1+U_1^{-2}U_2+U_1^{2}U_2+U_2^{-1}=0. \ee This can
be described by a homogeneous polynomial in a weighted projective
space. Indeed, we consider $\WP^2_{\la_1,\la_2,\la_3}(x_1,x_2,x_3)$
and introduce the following general gauge invariants \be
U_1=\frac{x_1^{\la_3/g_1}}{x_3^{\la_1/g_1}},\ \ \ \ \ \ \
U_2=\frac{x_2^{\la_3/g_2}}{x_3^{\la_2/g_2}} \label{U1U2} \ee where
\be g_1=gcd(\la_1,\la_3),\ \ \ \ \ \ \ g_2=gcd(\la_2,\la_3).
\label{gg} \ee The division by these common divisors is in order to
keep the ratios in (\ref{U1U2}) minimal. The geometry
(\ref{mgeometry}) may then be written \bea
0&=&x_1^{2\la_3/g_1}x_2^{\la_3/g_2}x_3^{2\la_1/g_1+\la_2/g_2}
  +x_2^{2\la_3/g_2}x_3^{4\la_1/g_1}\nn\\
&+&x_1^{4\la_3/g_1}x_2^{2\la_3/g_2}
  +x_1^{2\la_3/g_1}x_3^{2\la_1/g_1+2\la_2/g_2}
\eea where we have multiplied by
$x_1^{2\la_3/g_1}x_2^{\la_3/g_2}x_3^{2\la_1/g_1+\la_2/g_2}$.
Multiplying by an additional factor of $x_2^{2\la_3/g_2}$ leads to
\bea 0&=&x_1^{2\la_3/g_1}x_2^{3\la_3/g_2}x_3^{2\la_1/g_1+\la_2/g_2}
  +x_2^{4\la_3/g_2}x_3^{4\la_1/g_1}\nn\\
&+&x_1^{4\la_3/g_1}x_2^{4\la_3/g_2}
  +x_1^{2\la_3/g_1}x_2^{2\la_3/g_2}x_3^{2\la_1/g_1+2\la_2/g_2}.
\label{0xxx} \eea
\noindent{\bf Elliptic solution}
\newline
Now we introduce the parameters (or coordinates) \be
z_1=x_2^{\la_3/g_2}x_3^{\la_1/g_1},\ \ \ \ \ \ \
z_2=x_1^{\la_3/g_1}x_2^{\la_3/g_2},\ \ \ \ \ \ \
z_3=x_1^{\la_3/g_1}x_2^{\la_3/g_2}x_3^{\la_1/g_1+\la_2/g_2}
\label{vvv} \ee with weights \bea
(\mu_1,\mu_2,\mu_3)&=&((\la_1/g_1+\la_2/g_2)\la_3,
  (\la_1/g_1+\la_2/g_2)\la_3,2(\la_1/g_1+\la_2/g_2)\la_3)\nn\\
  &=&(\la_1/g_1+\la_2/g_2)\la_3\times(1,1,2).
\eea In terms of these parameters, (\ref{0xxx}) reads \be
0=z_1z_2z_3+z_1^4+z_2^4+z_3^2. \label{0vvv} \ee This corresponds to a
homogeneous polynomial of degree 4 in \be
\WP^2_{(\la_1/g_2+\la_2/g_2)\la_3,(\la_1/g_2+\la_2/g_2)\la_3,
  2(\la_1/g_2+\la_2/g_2)\la_3}=\WP^2_{1,1,2}.
\ee The mirror geometry given by (\ref{0vvv}) is seen to describe an
elliptic curve (since the degree satisfies $d=\mu_1+\mu_2+\mu_3$
in $\mathbf{WP}_{\mu_1,\mu_2,\mu_3}^2$).
\\
\noindent{\bf Non-elliptic solution}
\newline
A simple example is given by the  following  parametrization \be
U_1=\frac{x_1}{x_2},\ \ \ \ \ \ \ U_2=\frac{x_1x_2}{x_3^2},
\ee differing in form from (\ref{U1U2}). This results in a
polynomial constraint of the form \be
x_1x_2x_3^2+x_2^4+x_1^4+x_3^4=0 \label{pol111} \ee which corresponds
to a homogeneous polynomial in $\WP^2_{1,1,1}=\mathbf{CP}^2$. It is
ensured by construction that this satisfies the  mirror  constraint
which here reads \be (x_1x_2x_3^2)^4=x_2^4\times x_1^4\times
(x_3^4)^2. \ee The  algebraic  curve  (\ref{pol111}) is not elliptic.

Another non-elliptic curve follows from (\ref{U1U2}) when based on
$\WP^2_{1,2,3}$, for example, as the mirror is described by \be
x_1^3x_2^3x_3+x_1^{12}+x_2^6+x_3^4=0
\label{pol123} \ee with associated mirror constraint given by \be
(x_1^3x_2^3x_3)^4=x_1^{12}\times (x_2^6)^2\times x_3^4. \ee

\subsubsection{Folding procedure}

Having described the toric geometry of type IIB  superstring theory
in the presence of D7-branes and  its  mirror version, the next step
is  to show how  to convert the corresponding model into the
well-known  gauge theories living on the world volume of wrapped
D5-brane. This can  be done with the help of an outer automorphism
group action $\Gamma$ of the toric graphs of the Calabi-Yau
threefolds in the dual type IIB geometry. In this approach, $ADE$
toric graphs of ALE spaces  may be obtained, from type IIB
geometries  dual to F-theory on $ADE$  fourfolds,   by identifying
vertices  which are permuted by $\Gamma$.  Once this action has been
specified, one should solve the corresponding toric constraint
equations. It turns out that these can be derived easily from the
equation of the algebraic  curve  appearing in the mirror of toric
Calabi-Yau threefolds.

To understand how such  a surprising connection could be true, we
first consider the $A_1$ fourfolds in F-theory comaptification
corresponding to  ${\cal O}_{{\bf WP}^2_{1,1,2}} (-4)$. Indeed, let
us consider the  $\mathbb{Z}_2$ subgroup of $PSL(2,\mathbb{Z})$
acting as \be v_1\longleftrightarrow v_2, \ee that is, the vertices
$v_1$ and $v_2$ are in the same orbit of this action as they
transform as a doublet. Note that (\ref{toric1}) is invaraint under
this transfomartion. The folding procedure now amounts to
identifying these two vertices. In the mirror version, this folding
action is equivalent  to identifying the corresponding monomials as
\be U_1^{n_1^1}U_2^{n_1^2}=U_1^{n_2^1}U_2^{n_2^2}. \ee Using
(\ref{vertex}) and (\ref{U1U2}), this merely reduces to the
following simple constraint \be \label{constraint}
x_1^{\lambda_3}=x_3^{\lambda_1} \ee in the weighted space.
Implementing (\ref{constraint}) into (\ref{0xxx}), we get \be
0=x_2^{2\lambda_3/g_2}(x_2^{\lambda_3/g_2}x_3^{\lambda_2/g_2}
  +2x_2^{2\lambda_3/g_2}+x_3^{2\lambda_2/g_2}).
\label{superA1} \ee In terms of the new coordinates
$z_1=x_2^{\lambda_3/g_2}$ and $z_2=x_3^{\lambda_2/g_2}$, this
corresponds to \be
0=z_1z_2+z_1^2+z_2^2.
\ee This describes the deformation of an $A_1$ singularity of K3 and
is easily seen to result as the mirror of ${\cal
O}_{\mathbf{CP}^1}(-2)$ following from \bea y_1+y_2+y_3=0,
\\
\prod\limits_{i=1}^3y_i^{q_i}=1,
\eea where \be
q_i=(1,1,-2),
\label{vecA1} \ee cf. (\ref{mirror}).

We note that the Mori vector (\ref{vecA1}) can be obtained from the
Mori vector given in (\ref{vecwp2}). The above vertex
identification leads to the following reduction in the toric data of
the Calabi-Yau threefolds \be
(1,1,2, -4) \to 2(1,1,-2) \equiv (1,1,-2).
\ee The apperance of this vector is not surprising. It comes
originally from the $A_1$ geometry on which the F-theory is
compactified.  The key observation here is that our  procedure may
be regarded as the following topological change \be
{\cal O}_{\WP^2_{1,1,2}}(-4) \to {\cal O}_{\mathbf{CP}^1}(-2)\times
\mathbb{C}.
\ee
This topological change admits an interpretation  in terms
D-branes. On the left-hand side geometry, F-theory on an $A_1$
fourfold can be interpreted as type IIB superstring theory with
D7-branes having
world-volume $\mathbb{R}^4 \times \WP^2_{1,1,2}$ where the factor
$\WP^2_{1,1,2}$ is the compact space of ${\cal O}_{\WP^2_{1,1,2}}
(-4)$.  After  the folding procedure, this can be re-interpreted
in terms of D5-branes with world-volume $\mathbb{R}^4 \times
\mathbf{CP}^1$, where the factor $\mathbf{CP}^1$ is the compact part of
${\cal
O}_{\mathbf{CP}^1}(-2)$.
Moreover, we believe that the type IIB geometry becomes  a
{\em trivial} fibration of $ A_1$  ALE spaces over the complex plane.  In
this way, the D7-branes reduce to D5-branes wrapping the complex
curve $\mathbf{CP}^1$ in the blown-up $A_1$ singularity given by
(\ref{xyz2}). It has been shown that this quiver gauge theory has
$N=2$ supersymmetry, whereas the original gauge model living in the
D7-brane world-volume has $N=1$ supersymmetry. This requires that
the above $N=2$ supersymmetry should be broken down to $N=1$.
This reduction may be administered by the addition of a
tree level superpotential of the form
\be
W_{tree}=\frac{1}{i}\sum_jg_jtr_j \phi^j.
\ee
A similar mechanism has been
studied in the context of large-N duality in \cite{CIV}, see also
\cite{KL}. The introduction of  this superpotential   modifies
the geometry and leads to a {\em non-trivial} fibration of $A_1$ ALE spaces
over the complex plane. This modified fibration may be described by
\be
xy+z^2+W'(t)^2=0
\ee
where $t$ is the coordinate in the base, that is, in the complex plane.
It is noted that this new geometry
may undergo a geometric transition where the vanishing
$\mathbf{S}^2$ is replaced by an $\mathbf{S}^3$ while
the D5-branes are replaced by three-form fluxes.

\subsubsection{On mirror geometry}

Here, we would like to extend the above results to
higher-order Dynkin geometries. To start, it is recalled that a general
complex $p$-dimensional toric variety can be described by \be
V^p=\frac{C^{p+n}\setminus S}{C^{\star n}}
\label{pvar}
\ee where the $n$
$C^\star$ actions are given by \be C^{\star n}:\
z_i\mapsto\la^{q_i^a}z_i,\ \ \ \ \ \ \ i=1,\dots,p+n;\ \ \ \ \ \ \
   a=1,\dots,n.
\ee Requiring this to correspond to a Calabi-Yau manifold imposes the
conditions \be \sum_{i=1}^{p+n}q_i^a=0,\ \ \ \ \ \ \ a=1,\dots,n.
\label{CYgen} \ee A toric vertex realization reads \be
\sum_{i=1}^{p+n}q_i^av_i=0,\ \ \ \ \ \ \ a=1,\dots,n
\label{toricgen} \ee where the vertices $v_i$ are of dimension $p$.
In the case of a  toric Calabi-Yau manifold, we may choose the vertices as
\be
v_i=(m_i,1)
\label{vn} \ee as they implement the Calabi-Yau conditions in the sense that
\be
0=\sum_{i=1}^{p+n}q_i^av_i=(\sum_{i=1}^{p+n}q_i^am_i,\sum_{i=1}^{p+n}q_i^a)
  =(\sum_{i=1}^{p+n}q_i^am_i,0).
\label{0nq} \ee The $j$-th coordinate of $m_i$ is written $m_{ij}$
where $j=1,\dots,p-1$. The mirror manifold is given as a solution to
\bea
\sum_{i=1}^{p+n}b_iy_i&=&0\nn\\
\prod_{i=1}^{p+n}y_i^{q_i^a}&=&1,\ \ \ \ \ \ \ a=1,\dots,n.
\label{mmgen}
\eea
To solve these constraints,  one may  introduce
$p-1$ gauge invariants $U_j$ and write \be
y_i=\prod_{j=1}^{p-1}U_j^{m_{ij}}.
\label{yU} \ee This automatically solves the constraint equation in
(\ref{mmgen}) since \be
\prod_{i=1}^{p+n}\left(\prod_{j=1}^{p-1}U_j^{m_{ij}}\right)^{q_i^a}
  =\prod_{j=1}^{p-1}U_j^{\sum_{i=1}^{p+n}m_{ij}q_i^a}=1
\ee due to (\ref{0nq}). The mirror manifold is then given by \be
0=\sum_{i=1}^{p+n}b_iy_i=\sum_{i=1}^{p+n}b_i\prod_{j=1}^{p-1}U_j^{m_{ij}}.
\label{0aU} \ee
Assuming that this may be described by a homogeneous
polynomial in a weighted projective space, we introduce
$\WP^{p-1}_{\la_1,\dots,\la_{p}}(x_1,\dots,x_{p})$ and write \be
U_1=\frac{x_1^{\la_{p}/g_1}}{x_{p}^{\la_1/g_1}},\ \ \ \ \ \ \
U_2=\frac{x_2^{\la_{p}/g_2}}{x_{p}^{\la_2/g_2}},\ \ \ \ \ \dots,\ \ \ \ \ \
\
U_{p-1}=\frac{x_{p-1}^{\la_{p}/g_{p-1}}}{x_{p}^{\la_{p-1}/g_{p-1}}}
\label{UUU} \ee where \be
g_j=gcd(\la_j,\la_{p}),\ \ \ \ \ \ \ j=1,\dots,p-1.
\label{gj} \ee The mirror manifold may then be described by \be
0=\sum_{i=1}^{p+n}b_i
\left(\frac{x_1^{\la_{p}/g_1}}{x_{p}^{\la_1/g_1}}\right)^{m_{i1}}\times\dots
  \times\left(\frac{x_{p-1}^{\la_{p}/g_{p-1}}}{x_{p}^{\la_{p-1}/g_{p-1}}}
   \right)^{m_{ip-1}}.
\label{0axn} \ee In what follows, we will initially be interested in  the
case
$p=3$.

\subsubsection{$A_2$  quiver model}

Here we specialize to the case of $A_2$ quiver models in F-theory
compactification. The dual type IIB geometry  is given  by
a $U(1)^2$  linear sigma model   with 5 chiral  fields  with charges
\be
q=\left(\begin{array}{rrrrrr} 1&2&1&0&-4\\
0&1&2&1&-4\end{array}\right) \ee as Mori  matrix.   A simple vertex
realization is given by $v_i=(m_i,1)$ where \be m_1=(2,1),\ \ \
m_2=(-1,0),\ \ \ m_3=(0,-1),\ \ \
  m_4=(1,2),\ \ \ m_5=(0,0).
\ee
We will see that this geometry reduces  to an ALE space with $A_2$
singularity.  As before, we will base our analysis on mirror symmetry.
Indeed, the associated mirror manifold (\ref{0axn}) is thus defined
by \bea
0&=&b_1\left(\frac{x_1^{\la_3/g_1}}{x_3^{\la_1/g_1}}\right)^2
  \left(\frac{x_2^{\la_3/g_2}}{x_3^{\la_2/g_2}}\right)
  +b_2\left(\frac{x_1^{\la_3/g_1}}{x_3^{\la_1/g_1}}\right)^{-1}
  +b_3\left(\frac{x_2^{\la_3/g_2}}{x_3^{\la_2/g_2}}\right)^{-1}\nn\\
  &&+b_4\left(\frac{x_1^{\la_3/g_1}}{x_3^{\la_1/g_1}}\right)
  \left(\frac{x_2^{\la_3/g_2}}{x_3^{\la_2/g_2}}\right)^2+b_5.
\label{A2b} \eea For simplicity, we set $b_1=b_2=b_3=b_4=b_5=1$ in
the following. Multiplying (\ref{A2b}) by
$x_1^{\la_3/g_1}x_2^{\la_3/g_2}x_3^{4\la_1/g_1+4\la_2/g_2}$ leads to
\bea 0&=&x_1^{3\la_3/g_1}x_2^{2\la_3/g_2}x_3^{2\la_1/g_1+3\la_2/g_2}
  +x_2^{\la_3/g_2}x_3^{5\la_1/g_1+4\la_2/g_2}
  +x_1^{\la_3/g_1}x_3^{4\la_1/g_1+5\la_2/g_2}\nn\\
&&+x_1^{2\la_3/g_1}x_2^{3\la_3/g_2}x_3^{3\la_1/g_1+2\la_2/g_2}
  +x_1^{\la_3/g_1}x_2^{\la_3/g_2}x_3^{4\la_1/g_1+4\la_2/g_2}.
\label{examp3n2} \eea Let us introduce the coordinates \be
z_1=x_1^{\la_3/g_1}x_3^{\la_2/g_2},\ \ \ \ \ \ \
z_2=x_2^{\la_3/g_2}x_3^{\la_1/g_1},\ \ \ \ \ \ \
z_3=x_3^{\la_1/g_1+\la_2/g_2} \ee with weights \be
(\mu_1,\mu_2,\mu_3)=(\la_1/g_1+\la_2/g_2)\la_3\times(1,1,1). \ee In
terms of these, (\ref{examp3n2}) reads \be
0=z_1^3z_2^2+z_2z_3^4+z_1z_3^4+z_1^2z_2^3 +z_1z_2z_3^3.
\label{mirrorexamp3n2} \ee This corresponds to a homogeneous
polynomial of degree 5 in $\WP^2_{1,1,1}$. The folding action of
interest here identifies the vertices $v_2$ and $v_5$, \be
v_2\longleftrightarrow v_5 \ee
In the mirror geometry, this action
simply corresponds to \be z_2=z_3. \ee Having determined the toric
constraint, we can now derive the equation defining the mirror
geometry. A simple computation yields \be
0=z_2^5+z_1z_2^4+z_1^2z_2^3+z_1^3z_2^2=\sum_{i=1}^4{z}_1^{i-1}z_2^{6-i}.
\label{A2} \ee In terms of the new coordinates \be
y_i={z}_1^{i-1}z_2^{6-i}, \ee we  have \bea
\sum_{i=1}^4 y_i&=&0,\nonumber\\
\prod_{i=1}^4{y}_i^{q^a_i}&=&1,\ \ \ \;\; a=1,2
\eea
where
\bea
  q^1 &=&(1,-2,1,0),\nn\\
  q^2 &=&(0,1,-2,1).
  \eea
This coincides with  the mirror constraint equations  of an $A_2$
singularity
of K3.
This may alternatively be expressed through
\be
y_1y_3=  y_2^2,\ \ \ \ \ \,\,\,y_2y_4=y_3^2.
\ee

\subsubsection{More on the folding procedure}

The folding procedure identifying $v_k$ and $v_l$ imposes
the identification of $y_k$ with $y_l$ in the mirror geometry.
In terms of the gauge invariants (\ref{yU}), this implies that
\be
\prod_{j=1}^{p-1}U_j^{m_{kj}}=\prod_{j=1}^{p-1}U_j^{m_{lj}}.
\label{UU}
\ee
Assuming the association of the weighted projective space
$\WP^{p-1}_{\la_1,\dots,\la_{p}}(x_1,\dots,x_{p})$, cf. (\ref{UUU}),
the identification (\ref{UU}) may be expressed as
\be
\prod_{j=1}^{p-1}\left(\frac{x_j^{\lambda_p/g_j}}{x_p^{\lambda_j/g_j}}
  \right)^{m_{kj}-m_{lj}}=1.
\label{xx1}
\ee
To simplify our considerations, let us assume that the toric vertices of
the original model were chosen such that
\be
m_{1}=(1,0,\dots,0),\ \ \ \ \ \ \ m_{n+p}=(0,\dots,0).
\ee
In this case, (\ref{xx1}) reduces to
\be
1=U_{1}=\frac{x_{1}^{\lambda_p/g_{1}}}{x_p^{\lambda_{1}/g_{1}}}.
\ee
The new geometry obtained by the folding action identifying
$v_{1}$ and $v_{n+p}$ now follows from (\ref{0axn}). It reads
\be
0=\sum_{i=1}^{p+n}b_i
\left(\frac{x_2^{\la_{p}/g_2}}{x_{p}^{\la_2/g_2}}\right)^{m_{i2}}\times\dots
  \times\left(\frac{x_{p-1}^{\la_{p}/g_{p-1}}}{x_{p}^{\la_{p-1}/g_{p-1}}}
   \right)^{m_{ip-1}}
\label{0axn2} \ee
and describes (up to a simple rewriting) a polynomial equation
in the weighted projective space
$\WP^{p-2}_{\la_2,\dots,\la_{p}}(x_2,\dots,x_{p})$.

We see that $x_1$ and $m_{i1}$ no longer enter the game.
This suggests that the original geometry, whose mirror we just
found by folding, is described effectively by $n+p-1$ vertices.
Since $n$ is unaltered by the folding procedure, it follows that the
original toric variety is of dimension $p-1$, cf. (\ref{pvar}).
Thus, interpreting the new geometry
(\ref{0axn2}) as the mirror of a $(p-1)$-dimensional toric variety,
we should require that the set
\be
\{\hat{v}_i=(m_{i,2},\dots,m_{i,p-1},1);\ i=1,\dots,n+p-1\}
\ee
defines a toric vertex realization of a Calabi-Yau manifold whose
charge matrix satisfies
\bea
\sum_{i=1}^{p+n-1}\hat{q}_i^a&=&0,\ \ \ \ \ \ a=1,\dots,n,\nonumber\\
  \sum_{i=1}^{p+n-1}\hat{q}_i^a\hat{v}_i&=&0,\ \ \ \ \ \ a=1,\dots,n.
\eea
It would be interesting to pursue this general link further.

\section{Conclusion}

We have studied geometric engineering of $N=1$ $ADE$ quiver models.
In particular, we have considered a class of such models obtained by
the  compactification of F-theory on manifolds defined as elliptic
K3 surfaces fibered over certain $ADE$ 4-cycles. The latter are
constructed by solving the $ADE$ hyper-K\"ahler singularities. Our
main focus has been  on $A_r$ quiver models resulting when the base
space of the compactification fourfold of F-theory is built from
intersecting 4-cycles according to $A_r$ Dynkin graphs. The dual
type IIB superstring theory involves
   D7-branes on such cycles  embedded in toric Calabi-Yau threefolds. We 
have analyzed in
   some detail the cases of $A_1$ and $A_2$
   and found that they are
   linked to the $A_1$ and $A_2$ geometries of ALE spaces.
Our approach involves a particular geometric
   procedure referred to as folding. Using this, we have
   discussed how the
   physics  of F-theory on $ADE$ fourfolds  in the presence
   of D7-branes wrapping 4-cycles can be related to
    a scenario with D5-branes wrapping 2-cycles of ALE spaces.
\\[.4cm]
{\bf Acknowledgments.} AB would like to thank  C. Vafa for kind
hospitality  at Harvard University, where a part of this work is
done.  He would like also  to thank C. Gomez and E. H. Saidi for
discussions and scientific help.


\begin{thebibliography}{99}

\bibitem{KKV}
S. Katz, A. Klemm, C. Vafa, {\em  Geometric engineering of quantum
field theories}, Nucl. Phys. {\bf B 497} (1997) 173, hep-th/9609239.

\bibitem{KMV}
S. Katz, P. Mayr, C. Vafa, {\em Mirror symmetry and exact
  solution of $4d$ $N=2$ gauge theories I}, Adv. Theor. Math. Phys.
  {\bf 1} (1998) 53, hep-th/9706110.

\bibitem{BFS}
A. Belhaj, A.E. Fallah, E.H. Saidi, {\em On the non-simply
  mirror geometries in type II strings}, Class. Quant. Grav. {\bf 17}
  (2000) 515.

\bibitem{ABS2}
M. Ait Ben Haddou, A. Belhaj, E.H. Saidi, {\em  Geometric
Engineering of N=2 CFT$_{4}$s based on Indefinite Singularities:
Hyperbolic Case},
     Nucl. Phys. {\bf B 674} (2003) 593, hep-th/0307244.

\bibitem{ABS1}
  M. Ait Ben Haddou, A. Belhaj, E.H. Saidi {\em Classification of N=2
supersymmetric CFT$_{4}$s: Indefinite Series},
     J. Phys. {\bf A 38} (2005) 1793,  hep-th/0308005.

\bibitem{BS1}
A. Belhaj, E.H. Saidi, {\em Non Simply Laced Quiver gauge Theories
in Superstrings}, Afr. J. Math. Phys. {\bf 1} (2004) 29.

\bibitem{BDR}
  A. Belhaj, L.B. Drissi, J. Rasmussen, {\em On N=1 gauge models from
geometric
engineering in M-theory},
    Class. Quant. Grav. {\bf 20} (2003) 4973,  hep-th/0304019.

\bibitem{BS2}
     A. Belhaj, E.H.Saidi,  {\em  On HyperKahler Singularities},
     Mod. Phys. Lett. {\bf A 15}  (2000) 1767, hep-th/0007143.

\bibitem{Vafa1}
C. Vafa, {\em On $N=1$ Yang-Mills in four dimensions}, Adv. Theor.
Math. Phys. {\bf 2} (1998) 497, hep-th/9801139.

\bibitem{Vafa2} C. Vafa, {\em Evidence for F-theory}, Nucl. Phys.
{\bf B 469} (1996) 403, hep-th/9602022.

\bibitem{HV}
K. Hori, C. Vafa, {\em Mirror Symmetry}, hep-th/0002222.

\bibitem{HIV}
  K. Hori, A. Iqbal, C.  Vafa, {\em D-Branes And Mirror Symmetry},
hep-th/0005247.

\bibitem{A}
A. Belhaj, {\em  F-theory Duals of M-theory on G2 Manifolds from
Mirror Symmetry}, J. Phys. {\bf A 36} (2003) 4191, hep-th/0207208.

\bibitem{CIV}
F. Cachazo, K. Intriligator, C. Vafa, {\em A Large N Duality via a
Geometric Transition}, hep-th/0103067.

\bibitem{KL}
   K. Landsteiner, C.I. Lazaroiu, {\em  Geometric regularizations and dual
conifold transitions}, JHEP 0304 (2003) 028, hep-th/0303054.

\end{thebibliography}
\end{document}